\documentclass[a4paper,floatfix,rmp,showkeys,superscriptaddress]{revtex4}
\usepackage{amsmath,amssymb,amsfonts}            
\usepackage[latin1]{inputenc}
\usepackage{graphicx}
\usepackage{graphics}                 
\usepackage{color}                    
\usepackage{hyperref}      
\usepackage{url}
\definecolor{MyGray}{rgb}{0.93,0.93,0.93}
\makeatletter\newenvironment{graybox}{%
   \noindent\begin{lrbox}{\@tempboxa}\begin{minipage}{.985\columnwidth}}{\end{minipage}\end{lrbox}%
   \colorbox{MyGray}{\usebox{\@tempboxa}}
}\makeatother
\begin{document}

\title{In silico transitions to multicellularity}

\providecommand{\ICREA}{ICREA-Complex Systems  Lab, Universitat Pompeu
  Fabra   (GRIB),   Dr    Aiguader   80,   08003   Barcelona,   Spain}
\providecommand{\IBE}{Institut de Biologia Evolutiva, CSIC-UPF, 
Pg Maritim de la Barceloneta 37, 08003 Barcelona, Spain}
\providecommand{\SFI}{Santa Fe  Institute, 1399 Hyde  Park Road, Santa
  Fe  NM   87501,  USA}

\author{Ricard V. Sol\'e
\footnote{Corresponding author}}
\affiliation{\ICREA}
\affiliation{\IBE}
\affiliation{\SFI}   
\author{Salva Duran-Nebreda}
\affiliation{\ICREA}
\affiliation{\IBE}

\vspace{0.4 cm}
\begin{abstract}
\vspace{0.2 cm}  
The emergence  of multicellularity and developmental 
programs are among  the  major  problems  of  evolutionary biology.  Traditionally, 
research in this area has been based on the combination of data analysis and experimental work on one hand 
and theoretical approximations on the other. A third possibility is provided by computer simulation models, 
which allow to both simulate reality and explore alternative possibilities. These in silico models offer 
a powerful window to the possible and the actual by means of modeling how virtual cells and 
groups of cells can evolve complex interactions beyond a set of isolated entities. Here we 
present several examples of such models, each one illustrating the potential for artificial modeling 
of the transition to multicellularity.
\end{abstract}


\maketitle

\tableofcontents

\section{The physics of multicellularity}

The transition to multicellularity is tied to the emergence of interactions 
among previously isolated cells. As a problem in complexity (Anderson 1972) we could say that 
a multicellular system defines a level of organization whose global properties cannot 
be reduced to the properties of the individual units. This statement is relevant for many 
reasons. First, because the presence of an evolutionary innovation necessarily requires the 
cooperation between previously unrelated units (Schuster 1996). Once such 
interactions are in place, a network of connected elements needs to be considered 
in order to understand what is now at work. In early phases predating the transition to 
complex multicellular life forms, the network involved cell-cell as well as cell-substrate 
interactions. 

In dynamical terms, the transition required the emergence of cooperation among 
elements, which share a common space where they relate to each other and respond 
to environmental changes in a concerted manner.  Most theoretical and computational approaches to this problem do not take into account 
the fact that these systems are formed by physical objects and it might seem not so relevant when dealing 
with the generic mechanisms associated to cooperative dynamics. As usual, the level of detail that is used in a model 
scales with the type of question we wish to answer. If we search for general principles defining the 
appearance of cooperative aggregates, general models considering population dynamics and gene interactions 
are enough. But multicellularity also connects us with the first steps towards developmental programs and 
a previous step to other major innovations. Such innovations are always associated with novelties in the 
ways cells and tissues interact. The spatial arrangement of cells and the diversity potential provided by 
space and thus a explicit introduction of spatial degrees of freedom is essential.

Meaningful models of evolutionary dynamics of multicellular systems need to 
consider the role of generic physical mechanisms of morphogenesis that are not the result of complex regulatory processes. In 
this context, physical forces including gravity, adhesion or diffusion, and their generative potential, 
are considered (Newman and Comper 1990; Goodwin 1994). The interplay between these mechanisms allows for spontaneous pattern formation through segregation of cell types,  differential cell growth and mortality. Some of these generic, pattern-forming mechanisms likely predate the early history of both pre-cellular and multicellular life forms(Forgacs and Newman 2005; Sol\'e et al 
2007; Sol\'e 2009), along with others controlled by genetic circuits .

Using some of these mechanisms, an evolutionary model of form can be constructed. Moreover, since some of these mechanisms 
seem to strongly constrain the repertoire of potential structures that can be generated, they also 
offer a powerful framework to understand the origins of convergent designs (Alberch, 1980). In this context, as pointed out 
by John Tyler Bonner, simple explanations based on mathematical and computational models can help to grasp 
the principles of multicellular organization 
(Bonner 2001; Forgacs and Newman 2005). As noticed in (Newman and Bhat 2008) the interplay between 
physical constraints and genetic regulatory mechanisms has been traditionally 
overlooked in most studies, with few exceptions (see for example Eggenberger 1997; Coen et al 2004; Cummings 2006; 
Doursat 2008; Kaandorp et al. 2008). 

Although physics and embodiment are usually discussed in the context (or at the level of) organisms or tissues, 
there is another level of embodiment that requires attention: the external world, whose fluctuations and 
properties influence the repertoire of adaptations that can be available. Thus we could add other factors playing a role in the early stages 
of multicellularity, including the ecological context and the physics of the environment should also be taken 
into account. 

Here we present five examples of {\em in silico} models illustrating different aspects of 
the emergence of multicellularity. They involve different levels of complexity and address different questions, although all illustrate 
the generative potential of the role played by the basic physics that pervades the interactions among virtual cells.   

\vspace{0.25 cm}

\begin{graybox}
\vspace{0.5 cm}
{\bf Box 1: Modelling cell sorting under differential adhesion}
\vspace{0.5 cm}
\noindent

\vspace{0.25 cm}

The dynamics of cell sorting can be easily modelled by considering a simplified lattice model 
where the movement of cells is constrained by their local preferences. Specifically, let us consider a 
$L \times L$ lattice $\Omega$, i. e. a discrete set of sites: 
$$
\Omega = \{ (i,j) \in Z^2 \vert 1 \le i, j, \le L 1\}
$$
Each site $(i,i) \in \Omega$ is characterized by a "state", indicated as $S(i,j)$. 
This state can be $0$ if the site is empty and either $1$ or $2$ if occupied by 
cells. The two possible "cell" states indicate different cell types with different adhesion 
properties. Cells can be more or less prone to remain together and also might tend to either 
avoid or attach to the external medium. Since cell-cell (and cell-medium) interactions are necessarily local, 
(figure 1) a given cell can only interact with a maximum of eight nearest neighbors. 
If $J(S(i,j),S(k,l))$ indicates the energy associated to the interactions between the sites $(i,j)$ 
and $(k,l)$, a matrix of interaction coefficients will be defined, namely
$$
{\bf  J} = 
\left( \begin{array}{ccc}
J(0,0) & J(0,1) & J(0,2) \\
J(1,0) & J(1,1) & J(1,2) \\
J(2,0) & J(2,1) & J(2,2) \end{array} \right)
$$ 
\vspace{0.25 cm}
\end{graybox}

\begin{graybox}
\vspace{0.25 cm}
which is obviously symmetric, i. e. $$J(a,b)=J(b,a)$$ and $$J(0,0)=0.$$

The model 
allows cells to move to a neighboring position by switching the two local states 
provided that the final state is more likely to happen, i. e. consistent with 
the optimization of cell preferences. This can be obtained using an energy function, 
defined as
$$
{\cal H} = \sum_{i,j} \sum_{k,l} \left (1 - \delta_{s(i,j)s(k,l)} \right ) J(s(i,j), s(k,l))
$$
where the sum is performed only over nearest pairs. Here $\delta_{mn}=1$ when $m=n$ and 
zero otherwise, and thus the term $1-\delta_{mn}$ just discards pairs of sites with 
identical states. 

Each step, we choose a random site, another neighboring site and compute the new 
energy ${\cal H}'$ and compare it to the original one ${\cal H}$. If the difference 
$$\Delta {\cal H} = {\cal H}' - {\cal H}$$ is positive, an increase in energy would occur 
and thus is discarded. Instead, when $$\Delta {\cal H}<0,$$the largest the difference 
the more likely is the change to happen. This is a simplified model and more sophisticated 
approaches were developed by Glazier and co-workers, involving "cells" composed of connected 
sites defining a specific cell, whose shapes can change in realistic ways (Graner and Glazier 1992; Glazier and Graner 1993).

\vspace{0.25 cm}
\end{graybox}

\begin{figure}[b]
\includegraphics[scale=0.55]{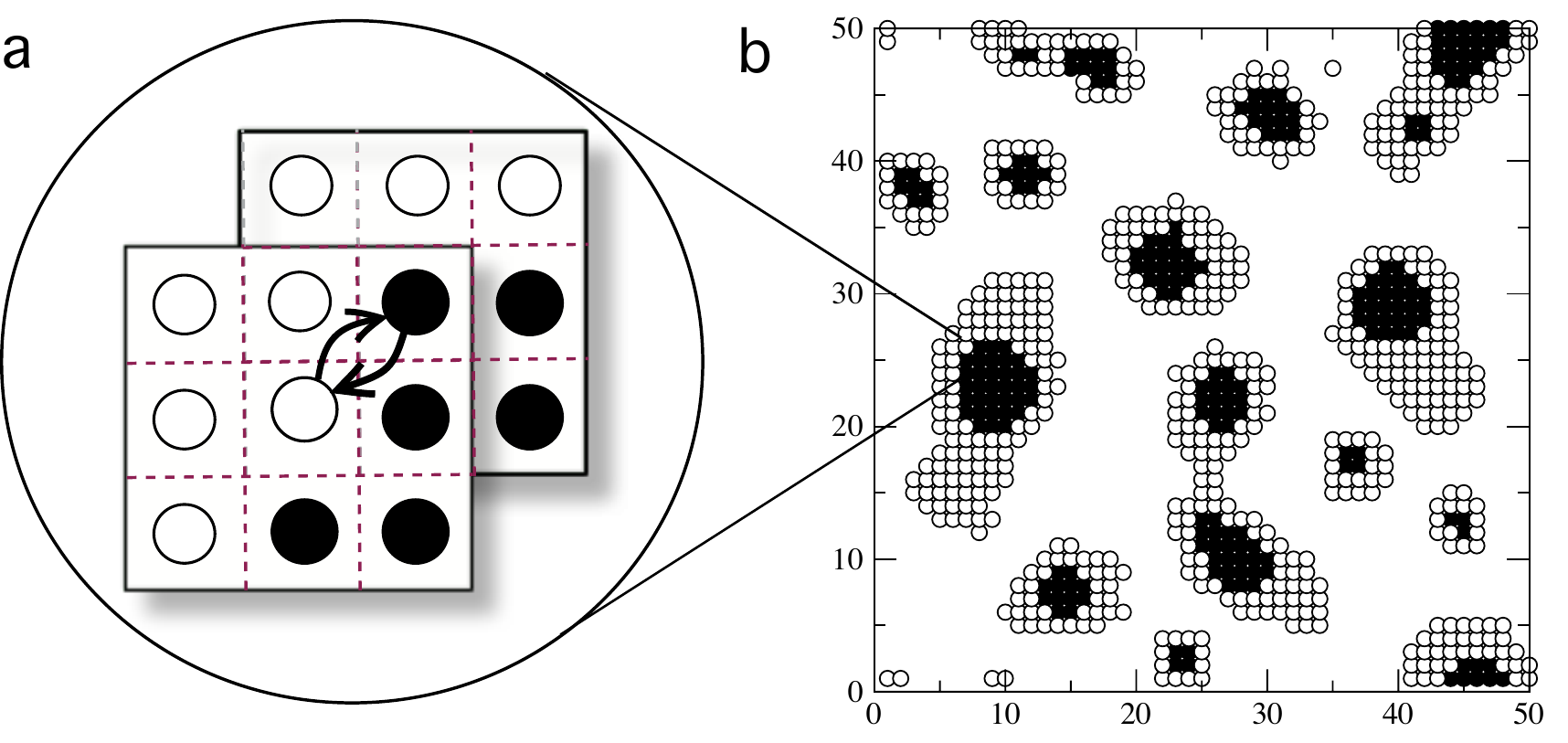}
\caption{Pattern formation through differential adhesion. Here a system composed by two types of cells (open and filled circles) 
are initially scattered at random on a two-dimensional lattice. Cells move and can flip their location with a neighboring cell 
(a) provided that the final energy is reduced (Box 1). Eventually, a stable arrangement of ordered cell assemblies  (b) 
is obtained. In this example, the adhesion parameters satisfy $J(1,2)>(J(1,1)+J(2,2))/2$ 
and $J(1,1)>J(2,2)$.}   
\end{figure}

\section{Evolving differential adhesion}

Our first example is a discrete model of evolution where digital "creatures" composed by many simplified 
cells interact through discrete adhesion forces. Hogeweg's model considers a population of 
of model organisms that is evolved using a {\em genetic algorithm} (Mitchell 1998; Forrest 1993) that allows 
to search over shape space under different selection pressures (Hogeweg 2000a, 2000b).
 Hogeweg's approach considers the growth and development of a simulated embryo. The model 
 description includes an internal boolean gene network (see Kauffman 1993), the evolution of 
 which leads to different adhesion among cells, cell division and death caused by stretching 
 and compression, cell migration and differentiation. 

Adhesion is introduced using very simplified but effective physical models (Graner and Glazier 1992; Glazier and Graner 1993; Sawill and Hogeweg 1997) and is one of the main players influencing the evolutionary dynamics of these virtual metazoans and their potential for diversification, consistent with the role played by development in the context of morphological radiations.  
Cell adhesion can easily promote the spatial organization of an initially disorganized, mixed group of cells. This is a very robust, repeatable and predictable mechanism of organization 
that amplifies initial disorder experienced by a mixed set of cells that move in space and aggregate with 
other cells under differential adhesion. The preferential adhesion mechanism is responsible for the sorting of cells 
in space, in a way that corresponds to the behavior of immiscible fluids (Forgacs and Newman 2005).

\begin{figure}[b]
\includegraphics[scale=0.55]{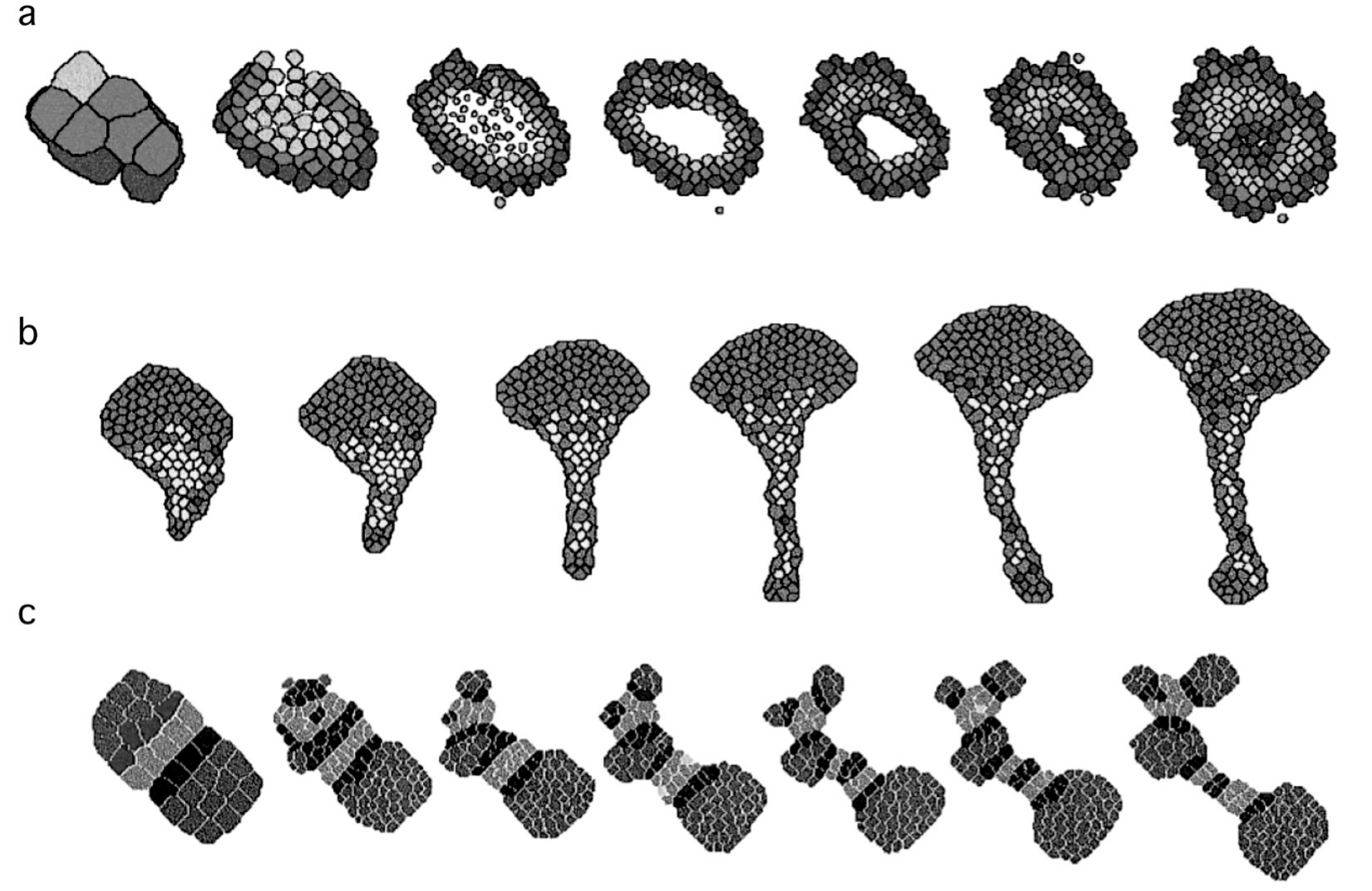}
\caption{Some examples of the developmental programs obtained in Hogeweg's in silico experiments 
involving the evolution of cell adhesion and differentiation (see Hogeweg 2000a). Differential cell adhesion leads
to cell migration and tissue remodeling, including intercalation of cells and cell layers. Here we show three 
different outcomes of the evolution process, where selection for maximal cell differentiation produces, as a side 
effect, morphogenetic processes and pattern formation. Among the developmental processes observable, we find: (a) cell 
migration and engulfing, (b) budding and elongation and (c) cell death and re-differentiation.}   
\end{figure}

As pointed out by Hogeweg (2000b) differential cell adhesion is regulated by the gene network affecting cell behavior 
and the communication between cells through cell-cell interactions. The model 
considers different types of fitness functions but the only strong pressure is directed 
to maximizing the diversity of cell types and thus there is no explicit search for 
special spatial arrangements or predefined developmental programs. Hogeweg's work revealed
 the existence of a neutral landscape of possible phenotypes that pervades the punctuated nature of transitions (Hogeweg 2000b; see also Fontana and Schuster 1998). Long periods of stasis are characterized by slow increases in fitness as small variations in phenotype are achieved. Selection for diverse gene expression patterns is used (see also Sol\'e et al. 2003), a choice that can be justified by an observed, well-known trend. The number of cell types is a good measure of complexity, which is known to increase through metazoan evolution (Carroll 2005). Increases in cell type number provide a high potential for further evolution of anatomical and functional complexity, essentially through division of labor and the formation of specialized tissues (Maynard-Smith and Szathm\'ary 1995). 
 
 Since the imposed selection pressure is rather generic, no special constraints are posed on the way genes interact and influence cell arrangements; no particular, predefined architectures and developmental plans are favored. The model is able to evolve complex forms, and in the process of evolving them, different remarkable changes take place. Complex shapes and some familiar ways of obtaining them (such as tissue engulfing, budding, etc) appeared and complex interactions between apoptosis or migration emerged. As pointed out in Hogeweg (2000a), morphogenesis itself emerges as a byproduct of optimization for cell diversity. It is worth noting that other works involving cell type richness as a fitness function favor the explosion of pattern forming motifs as soon as a threshold of genetic complexity is reached (Sol\'e et al. 2003).

\section{Multicellularity for free}

One of the most relevant and striking features of the transition to multicellularity is the fact that it took place multiple times in different lineages in the history of evolution (Abedin 2010; King 2004; Bonner 2001; Knoll 2011; Niklas and Newman 2013). For several authors this suggests that there is a certain component of inevitability in this process (Buss 1984; Grossberg and Strahmann 2007), that cooperation and specialization are such powerful innovations that convergent evolution into multicellularity ensued. In this vein, if Hogeweg's model proves that selecting for cell diversification can lead to co-select unexpected, emergent properties and behavior, this second model by Kaneko and colleagues (Kaneko and Yomo 1998) shows that even when there is no selection at work, cells can easily drift into multicellular phenotypes including differentiation and spatial patterning. 

As in the last example, the model starts with a single cell embryo that can develop into simple yet hallmark structures (see figure 4), just by allowing cell-cell and cell-environment interactions. The former are simulated by non-preferential adhesion between close cells and the latter by transport and processing of chemical species present in the virtual environment, which cells use to grow and divide. The internal workings of the cells are modeled by sets of coupled ordinary differential equations which describe autoregulatory transcription factor networks (see box 2). Remarkably, Kaneko's work shows that even if the internal networks, the initial state of the cell and the environment composition are randomized, a significant fraction of all possible cases develops consistently into virtual organisms with life-like features.

\begin{figure}[h]
\includegraphics[scale=0.6]{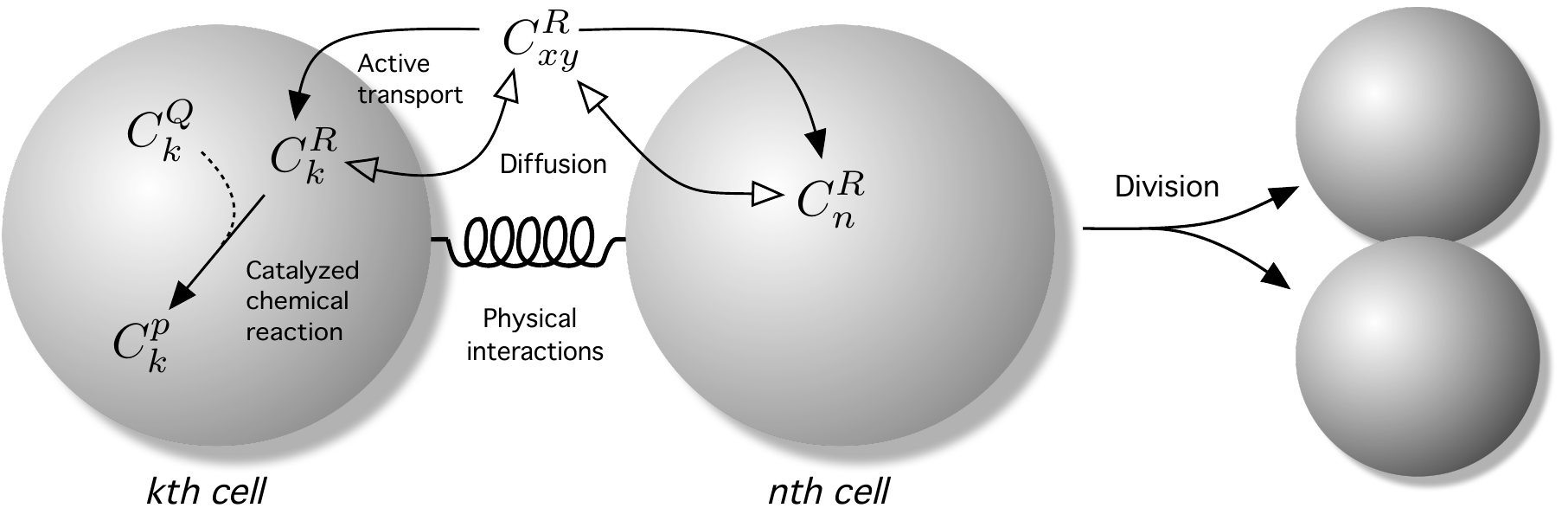}
\caption{Schematic representation of Kaneko's model, depicting the internal cell dynamics (catalyzed chemical reactions governed by 
sets of differential equations) as well as cell-cell interactions (diffusion and active transport of metabolites). The accumulation of key species up to a fixed threshold leads to division and stochastic allocation of the molecules between the new daughter cells. Redrawn from (Kaneko and Yomo 1999).}   
\end{figure}

\vspace{0.25 cm}

\begin{graybox}
\vspace{0.5 cm}
{\bf Box 2: Dynamical networks of differentiation}
\vspace{0.5 cm}
\noindent

\vspace{0.15 cm}
{\em Internal dynamics}
Cellular states are defined by the set of concentration of each species, namely ${C_{k}^1(t), ... , C_{k}^l(t)}$ for the concentration of the {\emph l} species in the  {\emph kth} cell at time  {\emph t}. A reaction matrix with notation $W(R, P, Q)$ is constructed, each position taking the value $1$ if there exists a reaction from chemical $R$ to chemical $P$ catalyzed by $Q$ and $0$ otherwise. The reaction matrix is generated randomly before each simulation and remains fixed during that simulation. Concentration change in this simple set of 3 species, namely Reactant  $R$, Product $P$ and Catalyzer  $Q$.  The equations read:

\vspace{0.25 cm}
\end{graybox}

\begin{graybox}
\vspace{0.25 cm}

$${\partial C_{k}^P \over \partial t} = W(R, P, Q)C_{k}^R (C_{k}^Q)^2$$
$${\partial C_{k}^R \over \partial t} = -W(R, P, Q)C_{k}^R (C_{k}^Q)^2$$

\vspace{0.15 cm}
{\em Cell-Environment dynamics}
The environment is simulated by a lattice of equal sized patches, each one of them characterized by ${C^1(x,y,t), ... , C^l(x,y,t)}$. Two different modes of material transfer are considered, passive diffusion and active transport. Diffusion takes place among neighbor finite elements and between occupied finite elements and the cells in them. It is assumed proportional to the difference of concentrations. Active transport, on the other hand, is considered to be only proportional to the concentration in the local finite element and modulated by the internal state of the cell. Thus the term accounting for diffusion and transport of the {\emph ith} chemical in the {\emph kth} cell is:
$$ {\partial C_{k}^i \over \partial t} = D_{e} \bigtriangledown^2 C^i(x,y) + D_{c}\sigma_{i}(C^i(x,y)-C_{k}^i)+T_{c}C^i(x,y)  \sum_{j=1}^l C_{k}^j $$

Where $D_{e}$ and $D_{c}$ stand for environment and cell diffusion coefficients respectively. The term $\sigma_{i}$ represents the membrane permeability to the {\emph ith} chemical and, as a simplification, can only take two values (0 or 1), fixed at random before the simulation. $T_{c}$ is the transport constant equal for all species and cells. The boundary conditions of the system are maintained to a high concentration of a source metabolite, which can be interpreted as feeding the tank where the simulated cells are placed with raw materials.

\vspace{0.25 cm}
{\em Cell division and death}
Division is considered to be the direct consequence of the accumulation of key, predefined chemicals inside the cell. A threshold function is defined, namely  $$\int_{t_{0}}^{t_{0}+\tau}C_{k}^ldt \geq D$$ 
as the criterion for cell division and splitting of resources between daughter cells, which are generated close to the position of the mother cell. 
Likewise, if the threshold condition  $$\sum_{i} C_{k}^i < S$$ is met, cells are 
removed from the simulation. 

\end{graybox}

\vspace{0.25 cm}

\begin{figure}[h]
\includegraphics[scale=0.62]{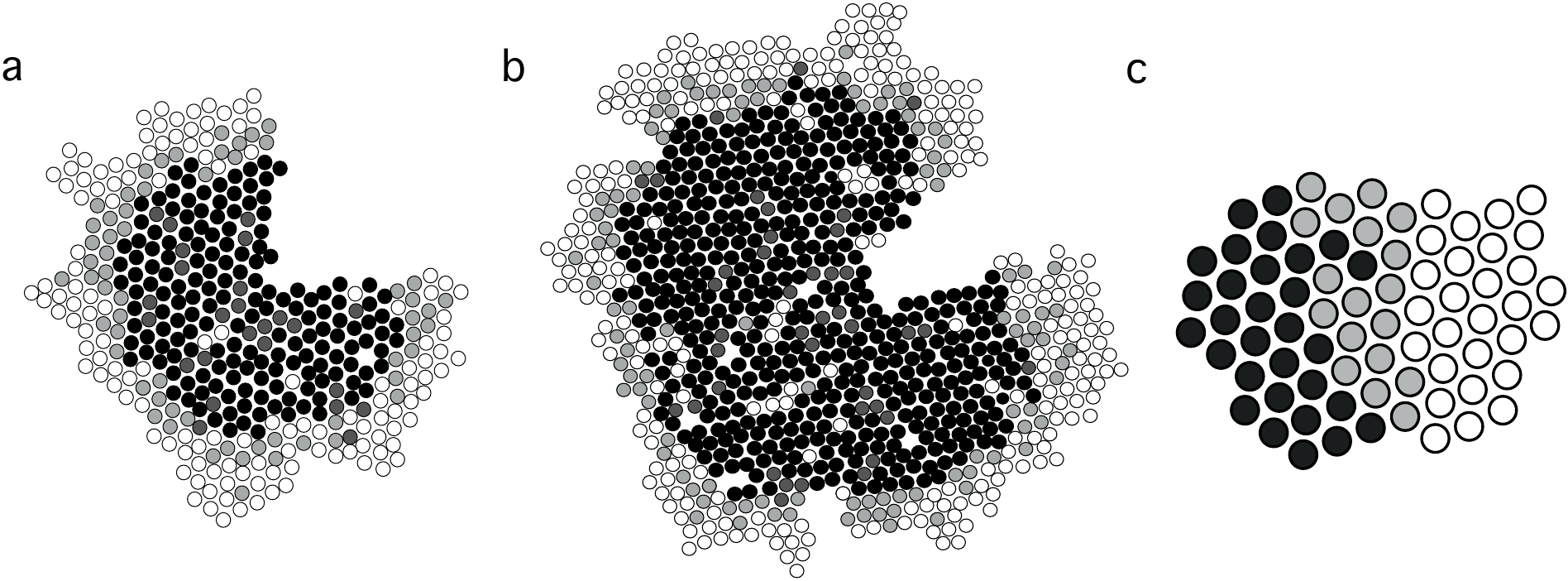}
\caption{Some of the multicellular organisms obtained by Kaneko's model (redrawn from Furusawa and Kaneko 1998; Furusawa and Kaneko 2003), the different shades of grey represent different cell types determined by mean metabolite concentration. Here we show the evolution of an organism after the removal of a quarter of the cluster (a), which leads to the regeneration of the previous ring pattern by means of differential cell growth and differentiation (b). Other natural occurring developmental forms are also observed, such as stripes (c).}   
\end{figure}

Kaneko and colleagues observed that from the subset of configurations displaying cell growth and intermediate connectivity (case study in Kaneko and Yomo 1999 contains 9 paths for a system of 32 chemical species), most led to periodic or quasi-periodic changes in metabolite concentrations inside the cells, analogous to natural cell cycles observed in both unicellular and multicellular organisms. Beyond a certain population threshold, synchronization and phase stabilization appeared among the cells in the same cluster, the first steps towards cooperation and collective action in cell societies. Further increase in the numbers within the ensemble enables the divergence of the mean chemical concentrations and the cycle itself (i.e. the differentiation of the cells). These new trajectories in the phase diagram were stabilized by mutually reinforcing metabolite exchanges between neighboring cells. 

This process of generation of variants from previously equal individuals, dubbed isologous diversification, provides a solid testbed to study both the community effect (Carnac and Gurdon 1997) and positional information (Wolpert 1969) theories. Moreover, after undergoing a first round of differentiation, cells could subsequently change into previously unavailable cell types, thus creating a tree-like hierarchical structure observed in natural developmental lineages.  Some of this virtual organisms displayed other key features of living systems like robustness to noise during development (Kaneko and Yomo 1999), resistance to "injury" through regeneration of spatial patterning (Furusawa and Kaneko 1998) and even life cycles exhibiting senile/proliferative stages (Furusawa and Kaneko 1998).

In conclusion, Kaneko's model demonstrates that even in the absence of evolution or selection, valid unicellular genotypes have the potential to create complex, emergent phenomena given that size thresholds can be surpassed. Whether this phenotypical changes suppose an increase in fitness is not of relevance here, but the feasibility to become multicellular and display potential task allocation as a side effect of cell-cell interactions. This results articulate a clear connection to Kauffman's work (Kauffman 1993) and the realization that some of the high order features observed in natural systems can arise not as a result of natural selection but the unavoidable properties of systems with high epistatic connectivity.  

\section{Evolving multicellular aggregates}

\begin{figure}[b]
\includegraphics[scale=0.65]{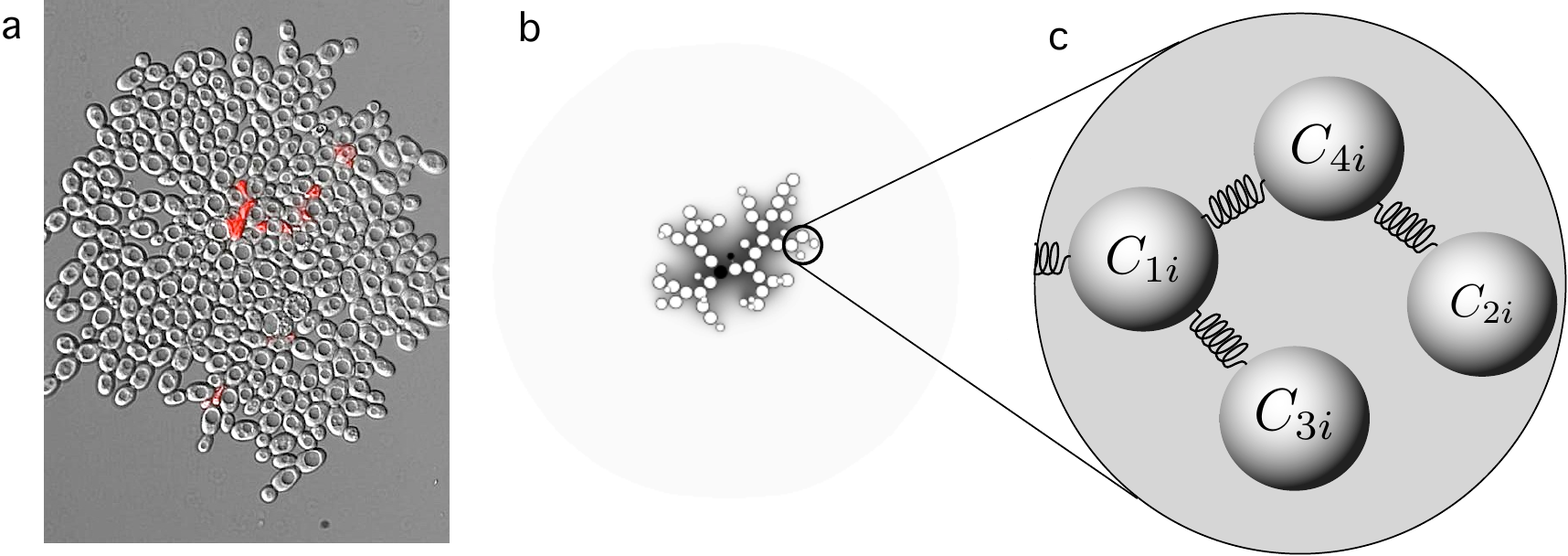}
\caption{The path from experimental results to in silico modeling. In (a) clusters of snowflake yeast with some apoptotic cells dyed in red in the center (from Ratcliff et al. 2012), breackage of a large cluster which generates juvenile propagules during the course of the simulation (b) and schematic representation of cell physics in the model (c) as described in box 3 (from Duran-Nebreda and Sol\'e 2014).}   
\end{figure}

Multicellularity has been a recurrent novelty in the story of life and some clues to its origins can be found (at least 
at the functional level) in living unicellular systems, such as bacteria or yeast. Many unicellular species display multicellular traits (Shapiro 1998; 
Bonner 2001) associated to the presence of signals that provide the source of coherent population responses. As a consequence of these 
responses (mainly to stress signals) multicellular aggregates can form and display some degree of specialization and/or differentiation.   
Following these observations, a very promising approach to study the feasibility of the transition to multicellularity is the use of artificial selection in natural systems. 

This strategy was put forward in a recent set of experiments (Ratcliff et al. 2012), in which the authors sequentially subcultured \emph{S. cerevisiae} cells with the fastest sedimentation in order to force the selection of cooperating aggregates. Yeast is a specially interesting candidate to explore the potential first steps due to the fact that it already presents some pre-adaptations thought to be relevant in this major transition in evolution (Maynard-Smith and Szathm\'ary 1995; Szathm\'ary 1994) and its biology and multicellular states are enough well described so that new emergent phenomena can be easily distinguished from them. Remarkably, after just 60 selection events -in a timescale much shorter than previously thought- the so called snowflake phenotype appeared consistently in all cultures. These are roughly spherical clusters of cells formed not by aggregation but by defective separation of cells after division. 

The clonal formation of the clusters ensures limited conflict of interest among the elements of the new multicellular individual, as discussed elsewhere (see Grossberg and Strahmann 2007; Bonner 2001; Michod 2000). Later on, the authors studied the role played by cellular fate in cluster reproduction. It was found that clusters did not reproduce through events associated to single cells but instead involved a group-level set of events that lead to the generation of a propagule. This was achieved through a division of labor in the form of the active control of apoptosis, which caused the asymmetrical splitting of the cluster once it reached a threshold size. 

In order to test alternative explanations to some of the phenomena observed -particularly the presence of a group-level reproduction and its relation to division of labor- and test other potential scenarios for the rise of multicellular ensembles, a simple embodied computational model was created 
(Duran-Nebreda and Sol\'e 2014). In it yeast cells are modeled as point elements in a bidimensional lattice which can fail to separate correctly after division, thus remaining attached by a spring. Cell's movement is modeled by a biased random-walk. No explicit genetic network is implemented, instead cells inherit the mother's parameters with small deviations. Two causes of cell death are tested, apoptosis as well as a simpler alternative based on the depletion of resources.

\vspace{0.25 cm}

\begin{graybox}
\vspace{0.5 cm}
{\bf Box 3: Evolving multicellularity under size-dependent selection}
\vspace{0.5 cm}
\noindent

\vspace{0.25 cm}

{\em Cell death caused by nutrient depletion}
To take into account this process, cells are placed in a bidimensional lattice that holds nutrient.  Cells have an energy value $M_{ni}$, a division threshold $M_{ni}^c$, a counter on the number of divisions $\Delta_{ni}$ and an attachment probability to daughter cells once they divide $p_{ni}$. 
Nutrient concentration change in the finite element $R_{ij}$  is given by the following equation:
$${\partial R_{ij} \over \partial t} = D \bigtriangledown^2 R_{ij} - \rho \Theta_{ij} R_{ij} $$
The Heaviside function $\Theta_{ij} $ is used to indicate the presence or absence of cells in that 
particular patch of the lattice (so $\Theta_{ij} =1$ if a cell is present and zero otherwise). 
The diffusion operator $\nabla^2 R_{ij}$ is numerically computed following a standard discretization form: 
$$
\nabla^2 R_{ij} = {\partial^2 R_{ij} \over \partial x^2} +  {\partial^2 R_{ij} \over \partial y^2} \approx  \left [ R_{ij} - {1\over4} \sum_{kl} R_{kl} \right]
$$
where $D$ accounts for the coefficient diffusion and $\rho$ the intake of nutrients from the culture medium. The energy change for $C_{ij}$ is: 
$$
{\partial M_{ij} \over \partial t} = \rho R_{ij} - \beta_c M_{ij} (1+\kappa \Delta_{ni})
$$
Where $\beta_c$ represents the maintenance costs. If the energy value of a particular cell reaches its division threshold, a new cell is created and the original energy value is split asymmetrically between the cells. Conversely, cells die if: $M_{ni} \leq \delta_c$, 
Where $\delta_c$ is the energy limit cells can withstand. 
\end{graybox}

\vspace{0.25 cm}

\begin{figure}[h]
\includegraphics[scale=0.85]{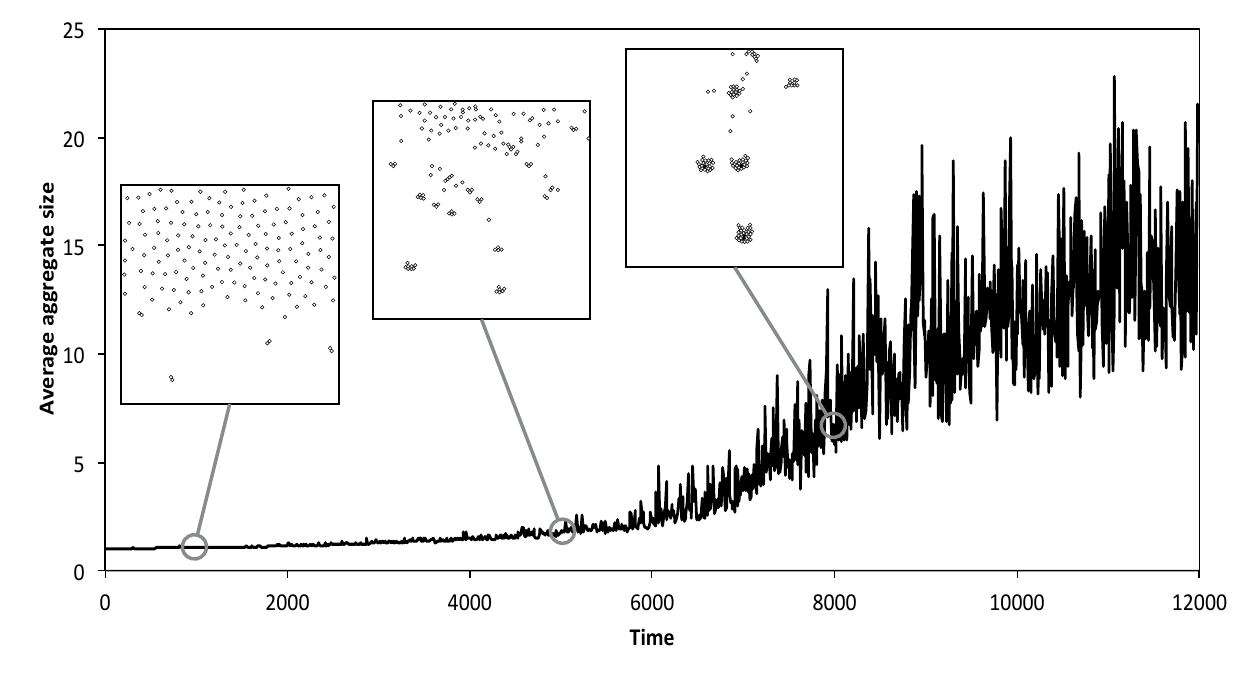}
\caption{\emph{In silico} evolution of aggregates under the nutrient depletion model (Duran-Nebreda and Sol\'e 2014). Average cluster size increases over time and stabilizes after $10^4$ simulation steps (mean for 10 replicate experiments). The maximum aggregate size is dependent on the nutrient concentration fixed at the beginning of the simulation (data not shown).}   
\end{figure}

The conclusions extracted from the simulations draw a slightly different picture on the possible first steps towards multicellularity. In it, physical constraints previously linked to a decrease in fitness for the aggregates -namely the added difficulty to attain enough nutrients to survive caused by limited diffusion in the core- are shown to work in favor of reproductive fitness, debilitating the cluster structure and promoting splitting after a certain size threshold is achieved.

\section{Physical forces and ecological scales}

An important component of a consistent theory of multicellularity, particularly in relation to the 
emergence of not just cooperation, but also developmental programs, requires considering community ecology 
in embodied models. Such models introduce physics and spatial interactions, and they naturally incorporate 
some selection forces, since the explicit physics introduce strong constraints to the potential forms and 
multicellular aggregates that can be obtained. If cell aggregates explicitly move, adhere to substrates, 
develop cooperation through cell-cell shared nutrients and resist external fluctuations, the evolution of 
adhesion and other variable features will occur under well-defined selection pressures. If these changes 
take place in a physical environment where available resources spread and are consumed, ecosystem-level 
processes might take place, some particular structures, such as budding, coherent multicellular shapes, differentiated aggregates 
or even life cycles could emerge. Moreover, potential changes in grazing efficiencies and the rise of predators 
can trigger arm races tied to changes in developmental programs, as it is likely the case for the 
transition between the Ediacaran and Cambrian biotas (Marshall 2006; Fedonkin 2007; Erwin and Valentine 2013).

\begin{figure}[h]

\includegraphics[scale=.65]{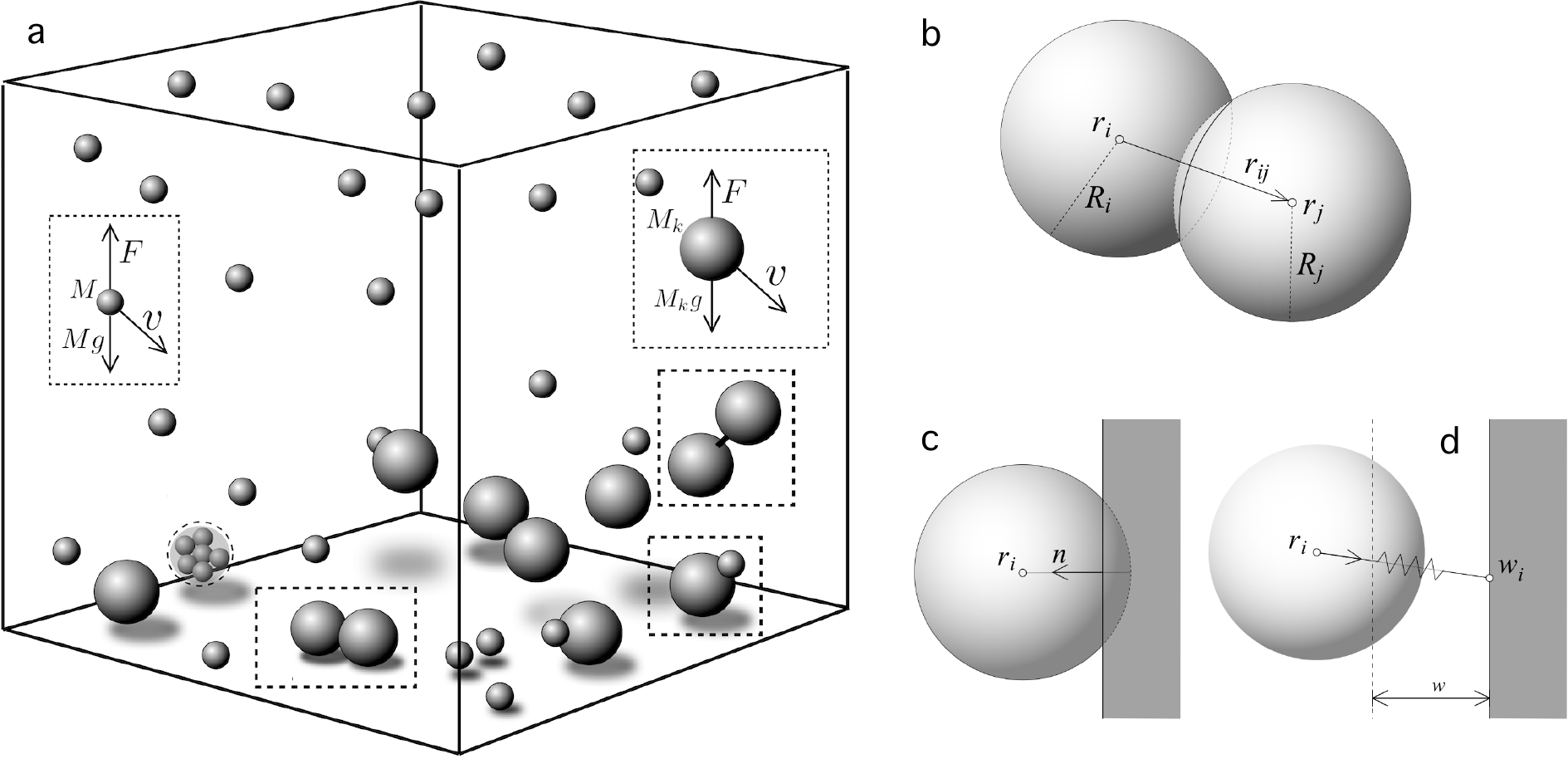}

\caption{Basic  scheme of  the components  of the CHIMERA  model (Sol\'e 
and Valverde 2013). The system (a) is confined within a  rigid cube.  Nutrient particles
  fall from the top layer experiencing physical forces. Cells also experience 
  the same forces, as described by Newton's laws (see Box 4). Additionally, both cells and particles 
  get degraded. Cells can interact with the boundaries of the system as well as 
  between them. Cells increase in mass every time they collide with a nutrient particle 
  if they have the right internal metabolism. We also display the mechanical forces acting between cells (b) 
  and the interactions between cells and the boundaries (c-d). In both cases, adhesion forces stabilize interactions 
  within some range, but interpenetration is forbidden.}
\label{chimera}  
\end{figure}

\begin{graybox}
\vspace{0.5 cm}
{\bf Box 4: Modeling ecology, physics and evolution}
\vspace{0.5 cm}
\noindent

\vspace{0.25 cm}

The CHIMERA model was build as an advanced framework introducing artificial 
cells as particles in a physical world where Newtonian forces, along with selection pressures, 
genes and metabolism are taken into account.

\vspace{0.25 cm}

{\em Cells and particles} Our starting point  is a population of single-cell organisms, 
where each cell in the initial  population is identical. 
Cells and particles are  simulated with rigid bodies moving within a fluid-like 
environment.  A cell  (particle) has spherical
geometry with radius $R_i$, mass $M_i$, spatial position ${\bf
  r}_i$ and  velocity $${\bf  v_i} = d{\bf  r}_i/dt .$$  The motion  of a
cell is described by the standard second law:
$$
M_i {d {\bf v}_i \over dt} = {\bf F}_i 
\label{eq_cell_force}
$$
Cell movement is obtained by numerical integration of the Newton equations. Cell velocity at
time $t + \Delta t $ is thus: $${\bf v}_i(t  + \Delta  t) = {\bf  v}_i (t  ) + {{\bf  F}_i/{M_i}} {\Delta t}$$ 
where $\Delta  t$ is the size  of the integration step, and the total 
force acting on $M_i$ will be:
$$
{\bf F}_i = {\bf  F}_e + {\bf F}_i^c + {\bf F}_i^n  + {\bf F}_i^l -
k_d  {\bf  v}_i +  M_i {\bf  g}
$$

applied  to any  cell  is  the  sum  of
environmental forces ${\bf F}_e$, the gravitational field ${\bf  g}$, the
collision force ${\bf F}_i^c$,  the cell-wall adhesion ${\bf F}_i^n$
and the cell-cell adhesion ${\bf F}_i^l$ term, along with the drag force associated to 
the medium, i. e. $-k_d  {\bf  v}_i $.

\vspace{0.25 cm}

\vspace{0.25 cm}
\end{graybox}

\begin{graybox}
\vspace{0.25 cm}

{\em Cell-substrate adhesion} Attachment   of   cells   to   surfaces  may   provide   a   favorable
micro-environment for cell aggregates to develop. If $D({\bf r}_i)$ indicates the 
cell-to-wall distance, when a cell with adhesion strength to the substrate $J_i^f > 0$ is closer than
the adhesion range  $\delta_w > 0$, i.e., $$D({\bf  x}_i) <  \delta_w ,$$ we attach 
a spring  connecting  the cell  ${\bf  x}_i$  with its projection  on the wall ${\bf  x}_i^w$ 
(Figure 6). Now, the wall spring exerts the following attraction force:

$$
{\bf F}_i^n  = - k_s \left ( ||{\bf r}_i- {\bf r}_i^w || - d_s \right ) {
  {{\bf r}_i- {\bf r}_i^w} \over { ||{{\bf r}_i- {\bf r}_i^w}||} }
\label{cellwall_force}
$$

where $d_s$  is the spring  equilibrium distance, $k_s$ is  the spring
constant and  ${\bf F}_n^i = 0$ when  the cell is not  attached to any
spring.  Existing  cell-wall springs can  be removed with  certain probability
$1- q(i)$ or when the spring length is above the maximal length, i.e.,
$$||{\bf  r}_i- {\bf  r}_i^w|| >  d_s^m .$$  As we  will see,  cells can
evolve cell-wall adhesion  $J_i^f$ in order to maximize  the intake of
nutrient particles.

\vspace{0.25 cm}

{\em Cell-cell adhesion} Cells can form  aggregates  by  attaching to  other cells.  Each cell  has an intrinsic 
probability $J_i^c$ to create a  new adhesion link.  Given  two close cells  
located at ${\bf r}_i$ and  ${\bf r}_j$,  we will set  an adhesion  string connecting
them with probability $(J_i^c + J_j^c )/2$.  The adhesion force to any
cell  is the sum  of forces  contributed by  all the  active cell-cell
adhesion springs:
$$
{\bf  F}_i^l =  - \sum_j k_l \left  (||{{\bf r}_i-  {\bf r}_j}||  - d_l
\right ) {{\bf r}_{ij} \over  ||{\bf r}_{ij}|| }
\label{cellcell_force}
$$
where  ${\bf r}_{ij} = {\bf r}_i- {\bf r}_j$, $d_l$ is  the spring  equilibrium  distance, and  $k_l$ is  the
spring  constant.   Adhesion  springs  break spontaneously  with  rate
$\delta \approx 0.001$ or when the spring is very large, i.e., 
$$||{\bf r}_i-  {\bf  r}_j||  >  d_l.$$

\end{graybox}

\vspace{0.25 cm}

In order to incorporate all these components, we need to build a complex simulation framework able to capture 
the essential physics, the presence of a population of interacting artificial agents and mechanisms of evolving 
the parameters that weight different metabolic and adhesion properties. Such type of model belongs to the 
tradition of so called artificial life approaches (Ray 1991; Langton 1991, 1995; Sipper 1995; Adami 1998) which involve the study of 
artificial life-like systems in artificial environments (along with a wet version associated with the construction of 
living systems using genetic engineering techniques). Computational models, which can reproduce realistic 
scenarios or completely ignore them, provide an ideal framework to explore the generative potential of an evolving set 
of rules allowing structures to emerge through time. Following this view, an embodied model of evolution, the so called 
CHIMERA model (Sol\'e and Valverde 2013) was introduced as a way of including newtonian physics, fluctuations, 
evolution and ecology in a unified fashion.

\begin{figure}[b]

\includegraphics[scale=.6]{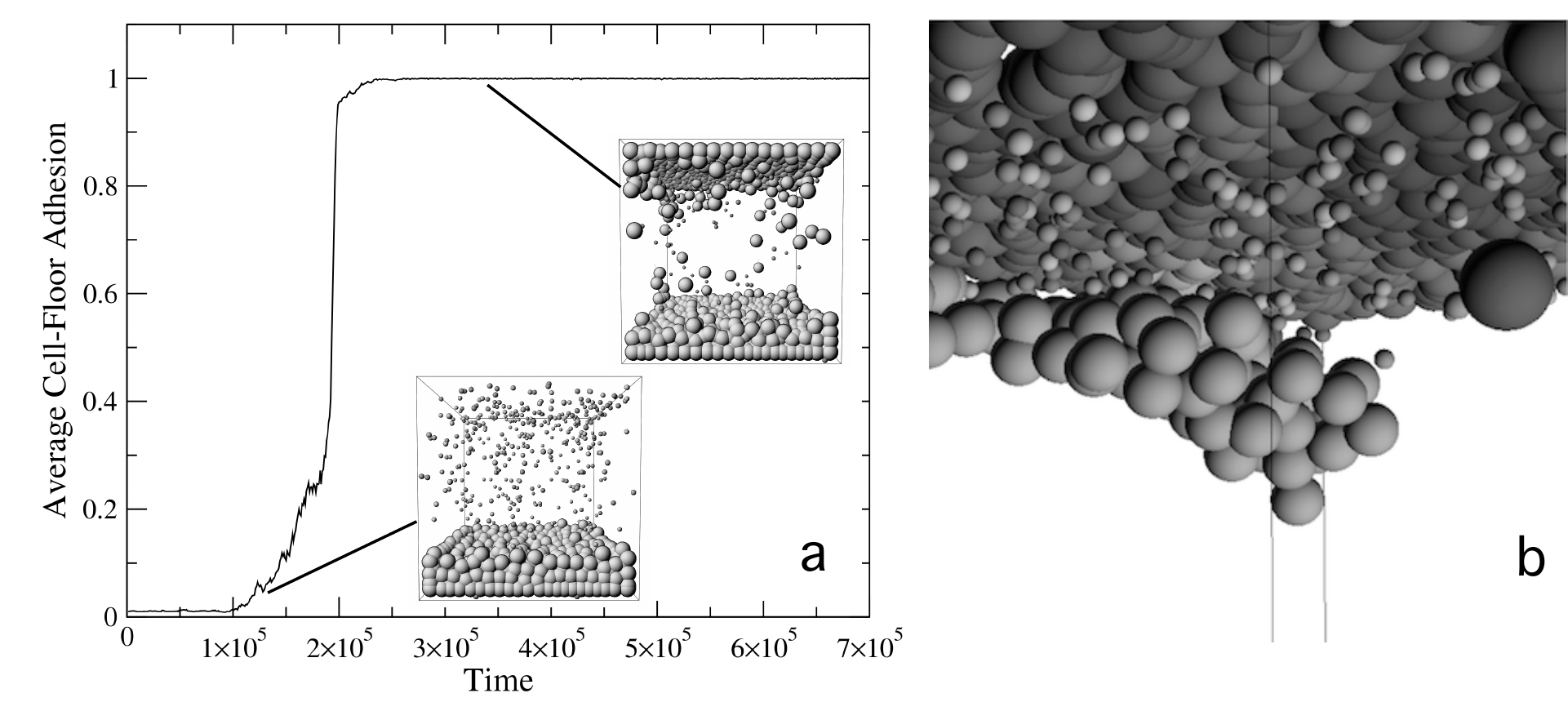}

\caption{(a) in the basic, newtonian CHIMERA model, the evolution of the system under 
enough nutrient levels drives the population from a few layers placed at the bottom to an 
inverted system where cells occupy and adhere to the upper wall. This occurs thanks to 
a broad distribution of efficiencies (i. e. evolved generalists) exploiting all resources, 
together with an increased adhesion between cells and specially between cells and surfaces. 
The main plot shows a fast growth in cell-floor adhesion towards its maximum value. Two snapshots of the 
system are also shown, before and after the transition.
In (b) an example of an evolved multicellular aggregate is shown. This was obtained by 
using a more detailed implementation of the cell-cell interactions that allow aggregates 
to emerge Sol\'e and Valverde 2013).}
\label{chimera}  
\end{figure}

The model was intended as an approach to the pre-Mendelian universe, which 
can be approached by studying the interplay between physical forces such as gravity, diffusion and adhesion and 
generic pattern-forming mechanisms. In its simplest version (Sol\'e and Valverde 2013) Chimera considers a cubic world involving a fluid-like 
medium with gravity and turbulence (figure 7) where an initial set of identical "cells" exploit one of a number of 
potential energy particles falling from the upper side of the cube. A set of rules is then used to evolve the system: 

\begin{enumerate}

\item
Movement: both particles and cells experience both a gravitational and a fluctuating velocity field (the later 
associated to turbulence). Particles are removed from the system with some probability. 

\item
Each cell carries a given set of internal parameters and variables: they have a given size and mass 
and they have a list of possible particle types that they can take and the efficiency of the grazing for 
each particle type. 

\item
Cells can attach to the surface of the walls with some probability. When they attach, a spring is used to properly define the physical interaction. Another adhesion probability is used for cell-cell adhesion. At the beginning all are set to zero. 

\item
If a cell interacts (collides) with a given particle, it ingests it if the efficiency for metabolizing that type of particle is non-zero. 
If taken, the mass of the particle gets transformed into mass of the cell. 

\item
Once a maximum cell size is reached, the cell splits into two daughter cells. Moreover, if the cell goes below a minimal 
value, it dies and it disintegrate. Detritus particles are also allowed to be part of the nutrient intake of cells. 

\item
Each time a cell divides, mutations can occur in the daughter cell. Metabolic efficiencies and 
adhesion rates can change. 

\end{enumerate}

The model is able to display complex forms of pre-multicellularity in terms of loose aggregates 
of cooperating cells. These aggregates evolve adhesion rates, both between cells but especially 
in relation to the physical substrate. Two major trends take place here. From the point of view of 
grazing efficiency, individual cells tend to become generalists: their metabolism changes in order to 
exploit all types of nutrient particles, although at the price of being less efficient. Secondly, in order 
to gather more particles, there is an advantage in providing a larger cell surface, which is possible 
provided that cells increase their attachment probability (adhesion force) and occupy available space 
on the lateral walls. This tendency, once started, is rapidly amplified and we can see (figure 8a) that 
eventually the cells "discover" the source of energy and occupy it. This is done by an increase of cell-floor 
adhesion but also increasing cell-cell adhesion and (when fluctuations are large) cooperation.

Eventually, the flow of particles is almost shut down except for cell mortality and the sedimentation of cells. 
Cell death creates detritus particles, which are also exploited with low efficiency.  But one a critical amount 
of detritus gets accumulated at the bottom, a new community of specialized detritivores emerges. 
This new ecosystem with two layers is stable over time and represents a dramatic example of how ecosystem engineering 
emerges: the feedbacks between organisms and environment trigger a control of the later by the first, 
with a deep reorganization of energy flows (Jones et al 1994, 1997; Hastings 2006; Erwin 2008; Erwin and 
Tweedt 2012). 

The outcome of the simulation reveals how a microscopic 
process of evolution associated with cell adhesion, combined with a community adaptation to the environment 
leads to a major innovation. Using simpler physics (Palsson and Othmer 2000; Palsson 2001; 
Ericson 2005; Sandersius and Newman 2008) where cells have volumes but their 
behavior is closer to point particles does not allow the formation of complex aggregates (Box 4). However, a more 
sophisticated and realistic definition of forces and spatial interactions allows the formation of multicellular aggregates (figure 8b) thus suggesting that much more can be obtained even under these simplified scenarios (Sol\'e and Valverde 2013).

\section{Combinatorial explosions and the Cambrian conundrum}

As a final example, we consider the problem of how complex and diverse spatial patterns 
of gene expression (and thus cell types) can emerge as a consequence of gene networks 
in development. Specifically, we consider an abstract model of pattern formation 
where a one-dimensional digital organism is composed of $C$ cells each carrying the same 
gene network. 

\vspace{0.25 cm}

\begin{graybox}
\vspace{0.5 cm}
{\bf Box 5: Gene networks in development}
\vspace{0.5 cm}
\noindent

\vspace{0.25 cm}

A simple, discrete model of pattern formation can be implemented by defining a set of $N=G+H$ genes 
interacting through a one-dimensional domain involving $C$ cells (Sol\'e et al., 2003). 
Here gene states will be Boolean: genes are either active ($1$) or inactive ($0$).  $G$ genes interacting within the cell, whose state at a given time $t$ 
will be indicated as $g_i^j(t)$, where $i=1, ..., G$ is gene number and $j=1, ..., C$ is cell 
number (ordered from anterior to posterior). The second group are generically labeled microhormones and 
actually involve (implicitly) some local mediators communicating neighboring cells. 

The state of these 
$H$ hormones will be indicated as $h_i^j(t)$. Hormones can receive inputs from any of the first $G$ 
units, but they can only make output to genes in other cells. Two matrices will be required in order 
to define the whole spectrum of links between different elements. These two matrices will be indicated 
by ${\bf W}=(W_{ik})$ and ${\bf C}=(C_{ik})$, defining interactions among the $G$ genes and between genes 
and hormones, respectively.

The basic set of equations of our gene network model read:
$$
g_i^j(t+1) = \Phi \left [ \sum_{l=1}^G W_{il}g_l^j(t) + 
\sum_{k=1}^H C_{ik} \delta \left( h_k^{j+1}(t) , h_k^{j-1}(t) \right) \right]
$$
where $\delta(x,y)=1$ if both $x=y=1$ and zero otherwise (i.e. an ``OR'' function). 
Similarly, hormones receive inputs only from inside the cell, 
$$
h_i^j(t+1)= \Phi \left [
\sum_{l=1}^G W_{il}g_l^j(t)
\right ]
$$
with additional, specific equations for the boundaries. 
The function $\Phi(x)$ is a threshold function, i. e. $\Phi(x)=1$ if $x>0$ and zero (inactive) otherwise. Specific 
equations are also defined for the two ends of the system.

Finally, we need to define an initial condition. The simplest choice that can be defined involves the 
activation of a single gene at the anterior ($j=1$) end of the digital organism. Specifically, we set $g_{i}^j(0)=h_{i}^j(0)=0$ for all $j=1, ..., N_c$, 
except $g_1^1(0)=1$. Thise choices corresponds to maternal signals confined to the chosen end. 
Such initial change will propagate to the rest of the tissue provided that the network allows it.

\end{graybox}

\vspace{0.25 cm}

The gene network includes both gene-gene interactions within the cell and between cells. 
In other words, we take into account regulatory interactions taking place within each cell together with 
cell communication through given signals (to be called hormones). These models have been 
extensively used since the early days of theoretical biology (Kauffman 1993) and and provide a simple 
way of approaching the problem of defining cell types and thus multicellular assemblies. 

In its simplest form, we can define a gene network in terms of a set of $n$ genes whose states $g_i$ are confined to two possibilities, 
namely $g_i \in \{ 0, 1 \}$. This binary approximation implies that genes are ON-OFF elements, which of course 
is a simplified picture. If the effect of gene $g_i$ on gene $g_j$ is given by a weight $W_{ij}$. If the gene 
interaction describes an activation or inhibition, $W_{ij}$ will be positive or negative, respectively. No interaction 
is given by $W_{ij}=0$. For simplicity, a discrete space of weights is used, namely $W_{ij} \in \{-1,0,+1\}$. In this 
way, a full exploration of the potential set of states can be performed. 

The state of a gene will change as a consequence of its interactions with other genes. 
This state is updated in discrete time units following:
$$
g_i^j(t+1) = \Phi \left [ \sum_{l=1}^n W_{il}g_l^j(t)  \right]
$$
which essentially tells us that the gene will become active or inactive if the global input acting on it 
is positive or negative, respectively. These networks can generate extremely simple (say, all genes inactive or 
active) or very complex dynamics (when chaotic changes occur). But for some ranges of connectivities, 
it leads to a rich diversity of stable states. If a cell type $T$ is identified as a string $\bf S_T$ of active 
and inactive genes, namely ${\bf S_T}=(g_1, ..., g_n)$ (Kauffman 1993) we have a potential of 
$2^n$ alternative strings. We can expand this formalism to take into account space, if multiple cells 
are also taken into account (Box 5). As shown in figure 9a, this is easily implemented and a spatial pattern can 
be described, for each gene, in terms of its expression level ($0$ or $1$) in different cells. 
A detailed analysis of this type of model reveals that some types of patterns are easily found (Sol\'e et al 2003; 
Munteanu and Sol\'e 2008; Tusscher and Hogeweg 2011). This is the case of regular stripe patterns, which is a rather common one. 

\begin{figure}
\includegraphics[scale=0.72]{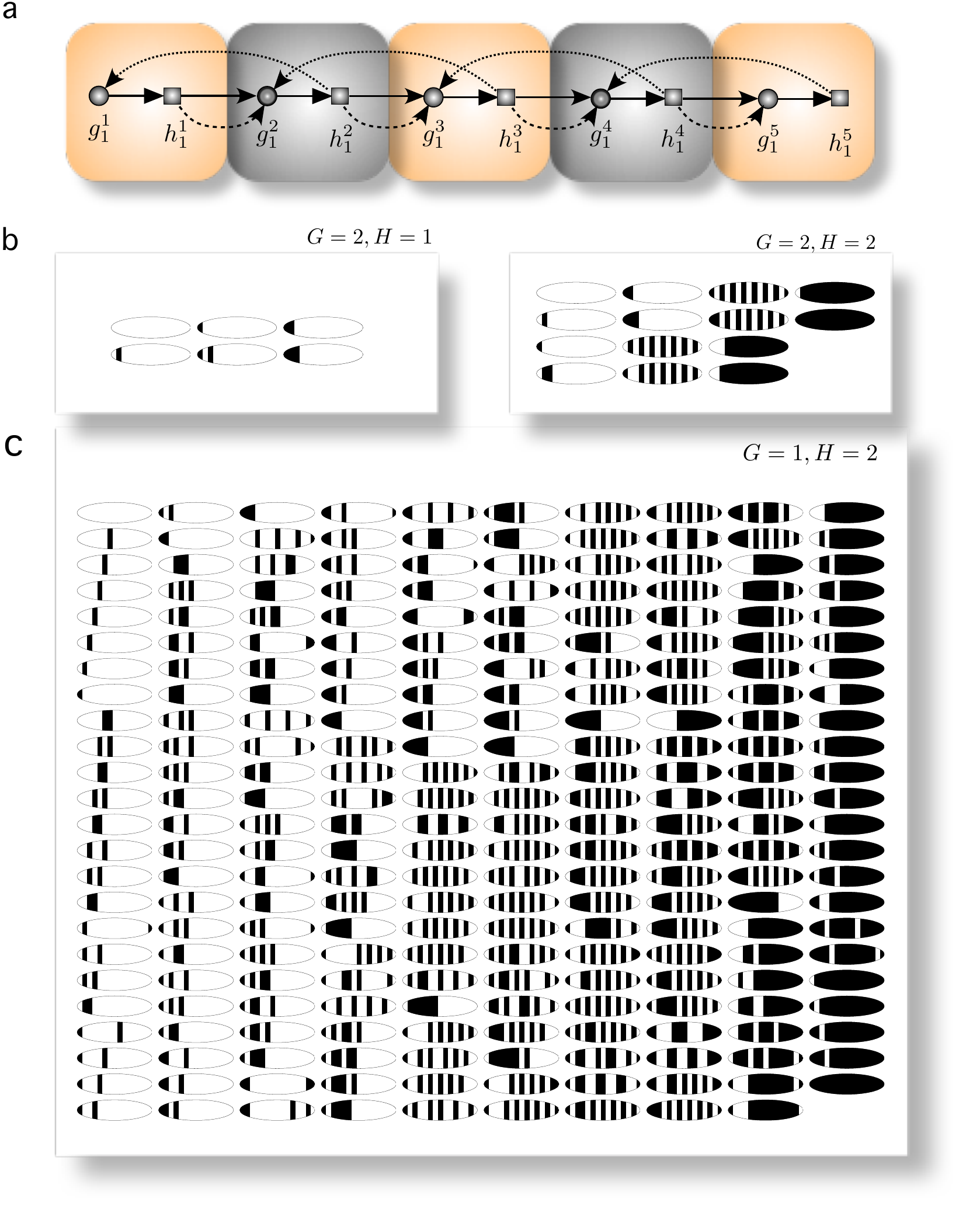}
\caption{Transitions from simple to complex (diverse) patterns in a minimal model of gene network model 
of biological pattern formation. The model considers a population of digital organisms composed 
by a linear chain of $C$ cells, each carrying the same genetic network (see Box 5). The complexity of this 
network, measured in terms of the number of genes $G$ associated to internal switches and the number of cell-cell signals 
$H$ determines the amount of patterns that can be achieved through evolution. 
In (a) dark and light 
cells indicate the high and low expression of one of the genes, respectively. In figures (b-c) the complete set of patterns generated 
in this way is shown for different sets of genes. For $H=2,G=1$ or $H=1,G=2$ 
only a few patterns are accessible (b). Once we have $H=G=2$, all possible patterns can be reached (c). }
\label{creature}  
\end{figure}

What is the generative potential of a given gene network complexity level? More precisely, if we start from a small 
network with $G=1$ genes and $H=1$ hormones and expand its number, what are the consequences? 
These questions are relevant when we consider the early events that shaped the gene networks affecting developmental 
processes in the history of multicellular life. Once again, thinking in the Cambrian explosion of life, 
several factors can be considered (Marshall 2006; Erwin and Valentine 2013). These include external, abiotic factors as well 
as internal ones and they are likely to interact among them. But it would be interesting to know, even under some basic abstract model, 
if some particular elements can play a key role in promoting sudden changes in the amount of achievable 
morphologies. In order to analyze this problem, we use the presence of a genotype-phenotype mapping $\Omega$, namely 
$$
\Omega : {\bf W} \longrightarrow \{ \Phi \}
$$
between the set of matrices $\bf W$ and the corresponding set of patterns $\{ \Phi \}$. In other words, for a given network $W_a$, the arrangement 
of ON and OFF genes defining a stable pattern $P_a^*$ can be written as $P_a^* = \Phi(W_a) \in \{ \Phi \}$. 

Using our basic model, it is possible to explore the space of potential phenotypes through single 
mutations in the genotype, as defined by the matrix of gene-gene interactions. Each step in the simulation, we evolve the 
system by changing single elements in the matrix. Patterns that are not stable are discarded and a change in the matrix 
is accepted if  the diversity of cells within the organism is increased (or at least remains the same). This movement in 
sequence space is known as an {\em adaptive walk} (Kauffman and Levin 1987). 

Pattern-forming gene networks display sharp thresholds affecting their  
combinatorial potential. For small numbers of elements, i. e. when $H+G<4$ and $H, G \le 2$ the range of possible 
spatial patterns is rather limited (figure 9b) but once the critical number $H=G=2$ is reached, all patterns become accessible 
(Sol\'e et al. 2003). This is a very interesting finding, since it provides a possible logical explanation for the 
rapid diversification of developmental paths when genetic complexity thresholds are crossed. Along with other influences, a small increase in regulatory complexity can 
account for a sudden jump in the achievable diversity of developmental pathways. 

Although these results are obtained from 
a toy model of regulatory interactions and ignore other pattern-forming factors,  such as tissue organization, 
morpho-dynamic processes or cell division, sorting and apoptosis, the basic conclusions are likely to be robust: 
a relatively small increase in underlying genomic complexity can lead to rich morphogenetic potential (Marshall 2006). 
In earlier models of evolution on fitness landscapes (Niklas 1994) high diversity is linked to 
the presence of multiple optima on a morphological landscape. If such optima are easily reached, a diverse range 
of structures is expected to be obtained. An interesting feature of the space of spatial patterns defined by 
the gene network model is that it displays neutrality: large, neutral networks percolate sequence space allowing for 
efficient exploration of the phenotype space. The structure of this pattern forming network space is actually very similar 
to other found in RNA folding (see Sol\'e et al 2003; Munteanu and Sol\'e 2008). This result tells us something 
important. As soon as we reach the critical threshold of network complexity, not only multiple patterns become accessible. 
The intrinsic evolvability of the system is also very high.

\section{Discussion and prospects}

In silico models of evolutionary change should be a natural component of our exploration of macroevolutionary patterns 
and the tempo and mode of evolutionary transitions. Despite their limitations, they offer, along with experimental dynamics using microbial populations (Lenski and Travisano 1994; Elena and Lenski 2003) what no other approach can: an opportunity to recreate potential scenarios and formulate hypothesis about how complexity developed over time. Here we have summarized the outcomes of different models of artificially evolved "organisms" or 
even communities where different types of transitions in complexity are achieved thanks to the (nonlinear) interplay 
between genetic and physical -and even ecological- components. Although they are all far from a realistic representation 
of developmental body plans or true communities, the previous results reveal a great generative potential implicit 
in the simple rules. In all cases, multicellular complexity undergoes increases or even jumps and some remarkable 
results can be highlighted. These include, for example, the emergence of some ontogenetic processes resulting 
from an evolutionary algorithm searching for diverse cell types.  Such processes typically incorporate cell-cell interactions 
that provide the capacity for tissue reorganization and shape changes together with cell differentiation. But even cellular and ecological 
scales can become related once evolving adhesion provides the exploratory potential for community-level processes 
to unfold. This connection between such disparate scales provides a novel way of re-considering the problem of hierarchies 
in evolution (Eldredge 1985; McShea 2001).

\begin{figure}
\includegraphics[scale=0.6]{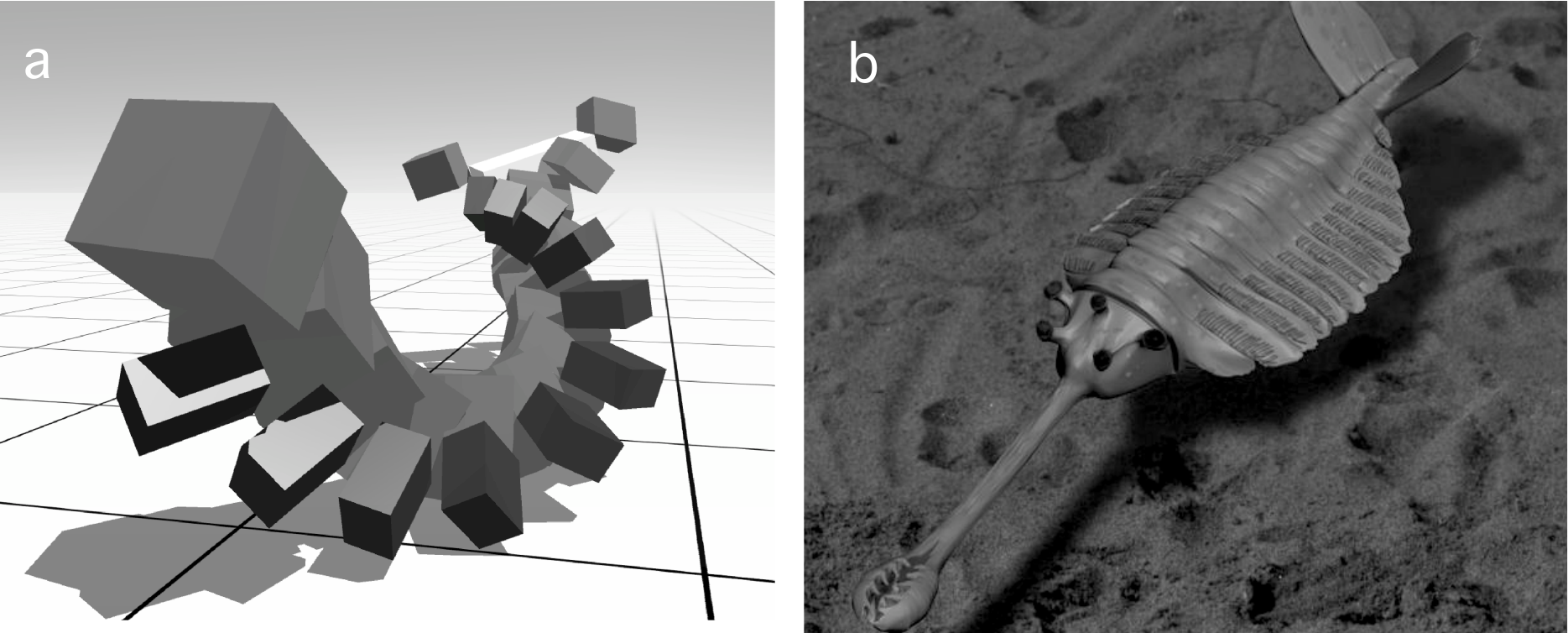}
\caption{Virtual creatures (a) can be obtained using physical models of interacting 
subparts defining a class of artificial life form and evolving under given selection constraints (rendering by 
Zach Winkler using the Stellar Alchemy program). Although they 
suggest that we can approach early life forms such as the Cambrian creatures like Opabinia (b), they strongly differ in the 
ways their complexity is generated (rendering by Nobu Tamura). }
\label{creature}  
\end{figure}

Some more sophisticated models have been created which are capable of evolving complex creatures with multiple connected components (figure 10a). 
These models involve a more or less detailed physical context, both in terms of the elements used to describe the 
virtual creature and the physics of the environment. Starting from Karl Sims work, these model involve a set of 
connected components (often parallelepipeds) linked to each other through strings (Sims 1994). The final outcome of this simulations often reminds us of some type of living creature. However, there is an essential difference: in the evolved artificial creatures there is no developmental program at work 
and thus there is no genotype-phenotype mapping. This is no minor drawback, since developmental programs are the essential 
component required to properly understand and model evolutionary paths. Over the last years, novel approaches 
to this problem incorporating some type of morphogenetic rules are being considered (Doursat 2008; Jin 2011). 

In the CHIMERA framework, Our artificial creatures are autogenic engineers (Jones 1994): they change their environment mainly via their own physical structures. The success of our model might be due to the complete set of key components that we allow to interact freely. By using space, diverse ecosystems can be build through spatial segregation of subpopulations. By allowing simple components of pattern formation or aggregate generation it is possible to introduce simple 
forms of cooperative dynamics. By embedding the virtual creatures within an ecosystem where physics plays a role, selection pressures 
restrict the repertoire of cellular aggregates that can be formed. 

Future work should address the potential for generating  complex 
structures perhaps similar to the Ediacaran fauna and test the role payed by both internal and external innovation triggers. The first 
includes for example the emergence of predators and the resulting arm races, which are known to be a major player in expanding 
morphological complexity. The second deals with extinctions caused by geological or astronomical shocks which deeply altered 
communities and whole ecosystems. The aftermath of the extinction provides a unique microscope to see different evolutionary 
processes in action. Such recovery patterns have been studied both from field data and modeling (Benton and Twitchett 2003; 
Erwin 1998, 2001; Sol\'e et al 2002; Chen and Benton 2012; Yedid et al 2012) and offer an additional test for studying how organismal 
and ecological complexity react to stress.

As a final point, it is worth mentioning that another avenue to address the dawn of multicellular systems is provided by synthetic biology (Benner and Sismour 2005; Sol\'e et al. 2007; Cheng and Lu 2012) which is considered by some researchers as the {\em wet} version of artificial life. By engineering unicellular systems, it is possible to 
build novel forms of cell-cell communication and thus create (and perhaps re-create) novel forms of 
multicellular assemblies, able to perform novel functions and even complex computations (Regot et al. 2011; Macia et al 2012; Chuang 2012). Given the potential offered by genetic engineering techniques to alter the 
logic of cell-cell exchanges, we have a unique opportunity of exploring the landscape of transitions from uni- 
to multicellular forms of organization. 
\vspace{0.25 cm}

{\bf Acknowledgements}

\vspace{0.25 cm}

We thank the members of the Complex Systems Lab for useful discussions. This work has been 
supported by grants of the James McDonnell Foundation, the Botin Foundation and by the Santa Fe Institute.

\vspace{0.25 cm}

\textbf{The present document is a chapter of the book titled \emph{Evolutionary transitions to multicellular life} (Springer) Edited by: A. Nedelcu and I. Ruiz-Trillo. The final publication is available at http://link.springer.com/}

\vspace{2.5 cm}

\noindent
{\bf References}

\vspace{0.25 cm}

\begin{enumerate}

\item
Abedin, M. and King, N. 2010. Diverse evolutionary paths to cell adhesion. Trends in Cell Biology 20, 734-742.

\item
Adami, C. 1998. Introduction to Artificial Life. Springer Verlag, New York.

\item
Alberch, P. 1980. Ontogenesis and morphological diversification. Amer. Zool. 20, 653-667.

\item
Anderson, P. 1972. More is different. Science 177, 393-396

\item
Benner, S.A. and Sismour, A. M. 2005. Synthetic biology. Nat. Rev. Genet. 6, 533-543.

\item
Benton M.J., and Twitchett R.J. 2003. How to kill (almost) all life:  the end-Permian extinction event. Trends Ecol. Evol. 18, 358-365.

\item
Bonner, J. T. 2001. First signals: the evolution of multicellular development. Princeton University Press. Princeton.

\item
Buss, L.W. 1987. The evolution of individuality. Princeton University Press. Princeton.

\item
Carnac, G. and Gurdon J.B. 1997. The community effect in Xenopus myogenesis is promoted by dorsalizing factors. Int. J. Dev. Biol. 41, 521-524.

\item
Carroll, S.B. 2005. Endless forms most beautiful: The New Science of Evo Devo and the Making of the Animal Kingdom. W. W. Norton and Co.

\item
Chen, Z.Q. and Benton, M.J. 2012. The timing and pattern of biotic recovery following the end-Permian mass extinction. 

\item
Cheng, A. and Lu, T.K. 2012. Synthetic Biology: An Emerging Engineering Discipline
Annu. Rev. Biomed. Eng.  14, 155-178.

\item
Chuang, J.S. 2012. Engineering multicellular traits in synthetic microbial populations. Curr. Opin. Chem. Biol. 16, 370-378.

\item
Coen, E., Rolland-Lagan, A.G., Matthews, M., Bangham, J. A. and Prusinkiewicz, P. 2004. The genetics of geometry. Proc. Natl. Acad. Sci. USA 101, 4728-4735.

\item
Cummings, F. 2006. On the origin of pattern and form in early metazoans. Int. J. Dev. Biol. 50, 193-208.

\item
Doursat, R. 2008. Organically grown architectures: creating decentralized, autonomous systems by embryomorphic engineering. In: Organic Computing, R. P. W\"rtz, ed.: pp. 167-200. Springer-Verlag.

\item
Duran-Nebreda, S. and Sol\'e, R.V. 2014. Emergence of multicellularity in a model of cell growth, death and aggregation under size-dependent selection. \emph{Submitted to} Proceedings of the Royal Society B.

\item
Eggenberger, P. 1997. Evolving morphologies of simulated 3d organisms based on differential gene expression. In. Fourth Europ. Conf. Artificial Life, P. Husbands and I. Harvey, editors. pp. 205-213. MIT Press.

\item
Eldredge, N. 1985. Unfinished Synthesis: Biological Hierarchies and Modern Evolutionary Thought. Oxford University Press. Oxford.

\item
Elena, S.F.and Lenski, R.E. 2003. Evolution experiments with microorganisms: the dynamics and genetic bases of adaptation. Nat. Rev. Genet. 4(6):457-69.

\item
Ericson, C. 2005. Real-Time Collision Detection. Morgan Kaufmann.

\item
Erwin, D. H. 1998. The end and the beginning: recoveries from mass extinctions. Trends Ecol. Evol. 13, 344-349.

\item
Erwin, D. H. 2001. Lessons from the past: biotic recoveries from mass extinction. 
Proc. Natl Acad. Sci. USA 98, 5399-5403.

\item
Erwin, D.H. and Tweedt, S. 2012. Ecological drivers of the Ediacaran-Cambrian diversification of Metazoa. Evol. Ecol. 26, 417-433

\item
Erwin, D.H. and Valentine, J. 2013. The Cambrian Explosion: The Construction of Animal Biodiversity. Roberts and Company.

\item
Fedonkin, M.A. 2007. The Rise of Animals: Evolution and Diversification of the Kingdom Animalia. JHU Press.

\item
Forgacs, G. and Newman, S. A. 2005. Biological physics of the developing embryo. Cambridge Univ. Press, Cambridge.

\item
Forrest, S. Genetic algorithms: Principles of natural selection applied to computation. Science 1993, 261, 872-878.

\item
Furusawa, C. and Kaneko, K. 1998. Emergence of multicellular organisms with Dynamic differentiation and spatial pattern. Artificial life 4, volume 1.

\item
Furusawa, C. and Kaneko, K. 2003. Robust development as a consequence of generated positional information. J. Theor. Biol. 224, 413-435.

\item
Fontana, W. and Schuster, P. 1998. Continuity in evolution: on the nature of transitions.
Science 29, 280(5368):1451-5.

\item
Goodwin, B. C. 1994. How the leopard got its spots. Princeton U. Press, Princeton.

\item
Glazier, J. A. and Graner, F. 1993. Simulation of the differential adhesion driven rearrangement of biological cells. Phys. Rev. E 47, 2128-2154.

\item
Graner, F. and Glazier, J.A. 1992. Simulation of biological cell sorting using a two-dimensional extended Potts model. Phys. Rev. Lett. 69, 2013-2016.

\item
Grossberg, R.K. and and Strahmann, R.R. 2007. The Evolution of Multicellularity: A Minor Major Transition? Annu. Rev. Ecol. Evol. Syst. 38, 621-654.

\item
Hastings, A. 2006. Ecosystem engineering in space and time. Ecol. Lett. 10, 153-164.

\item
Hogeweg, P. 2000. Evolving Mechanisms of Morphogenesis: on the Interplay between Differential Adhesion and Cell Differentiation. J. Theor. Biol. 203, 317-333 

\item
Hogeweg, P. 2000. Shapes in the shadow: Evolutionary dynamics of morphogenesis. Artif. Life, 6: 85-101.

\item
Jin, Y. 2011. Morphogenetic robotics: an emerging new field in developmental robotics. 
IEEE Trans. Syst. Man Cyber. 41,145-160. 

\item
Jones, C.G. , Lawton, J.M. and Shachak, M. 1994. Organisms as ecosystem engineers. OIKOS 69, 373-370. 

\item
Jones, C.G. , Lawton, J.M. and Shachak, M. 1997. Positive and negative effects of organisms as physical ecosystem engineers. Ecology 78, 1946-1957.

\item
Kaandorp, J.A., Blom, J.G., Verhoef, J., Filatov, M., Postma, M. and M\"uller, W.E.G. 2008. Modelling genetic regulation of growth and form in a branching sponge. Proc. R. Soc. B 275, 2569-2575.

\item
Kaneko, K. and Yomo, T. 1999. Isologous diversification for robust development of cell society. J. theor. Biol 199, 243-256.

\item
Kauffman, S.A. and Levin S. 1987. Towards a general theory of adaptive walks on rugged landscapes. J. Theor. Biol. 128, 11-45.

\item
Kaufmann, S.A. 1993. The Origins of Order: Self-Organization and Selection in Evolution. Oxford University Press.

\item
King, N. 2004. The unicellular ancestry of animal development. Developmental Cell. 7, 313-325.

\item
Knoll, A.H. 2011. The Multiple Origins of Complex Multicellularity. Annu. Rev. Earth Planet. Sci. 39, 217-239.

\item
Langton, C.G. 1991. Life at the Edge of Chaos. in {\em Artificial Life II}, Addison-Wesley.

\item
Langton,G.G. (ed) 1995. Artificial life: an overview. MIT Press, Cambridge MA

\item
Lenski, R.E. and Travisano, M. 1994. Dynamics of adaptation and diversification: a 10,000-generation experiment with bacterial populations. Proc Natl Acad Sci U S A. 91, 6808-14.

\item
Mac\'ia, J., Posas, F. and Sol\'e R.V. 2012. Distributed computation: the new wave of synthetic biology devices. Trends Biotechnology 30(6):342-9.

\item
Marshall, C.R. 2006. Explaining the Cambrian "Explosion" of animals. Annu. Rev. Earth Planet. Sci. 34, 355-384.

\item
Maynard Smith, J. and Szathm\'ary, E. 1995. The major transitions in evolution. Oxford University Press. Oxford.

\item
Michod, R. E. 2000. Darwinian Dynamics: Evolutionary Transitions in Fitness and Individuality. Princeton University Press. Princeton.

\item
Mitchell, M. 1998. An Introduction to Genetic Algorithms. Bradford Books.

\item
Munteanu, A. and Sol\'e, R.V. 2008. Neutrality and robustness in evo-devo: emergence of lateral inhibition. PLoS 
Comput. Biol. 4, e1000226.

\item
McShea, D.W. 2001. The hierarchical structure of organisms: a scale and documentation of a trend in the maximum. Paleobiology 27, 405-423.

\item
Newman, S.A. and Baht, R. 2008. Dynamical patterning modules: physico-genetic determinants of morphological development and evolution. Phys. Biol. 5, 015008.

\item
Newman, S.A. and Comper, W.D. 1990.'Generic' physical mechanisms of morphogenesis and pattern formation.Development, 110(1):1-18.

\item
Niklas, K.J. 1994. Morphological evolution through complex domains of fitness.
Proc. Natl. Acad. Sci. USA 91, 6772-79.

\item
Niklas, K.J. and Newman, S.A. 2013 The origins of multicellular organisms. Evol Dev 15:41-52

\item
Palsson, E. 2001. A Three-dimensional model of cell movement in multicellular systems. Future Gener. Comp. Syst. 17, 835-852.

\item
Palsson, E. and Othmer, H.G. 2000. A model for individual and collective cell movement in Distyostlium discoideum. Proc. Natl. Acad. Sci. USA 97, 10448-10453.

\item
Ratcliff, W.C., Denison, R.F., Borrello, M. and Travisano, M. 2012. Experimental evolution of multicellularity. Proc. Natl. Acad. Sci USA. 109, 1595-1600.

\item
Ray, T.S. 1991. An approach to the synthesis of life. In: Langton, C., Taylor, C. and Farmer, D. (eds) Artificial Life II. pp. 371-408.

\item
Regot, S., Macia. J., Conde. N., Furukawa. K., Kjell\'en. J., Peeters. T., Hohmann. S., de Nadal. E., Posas. F. and Sol\'e. R. 2011. Distributed biological computation with multicellular engineered networks. Nature 469(7329):207-11.

\item
Sandersius, S.A. and Newman, T.J. 2008. Modelling cell rheology  with the subcellular element model. Physical Biology 5, 015002.

\item
Savill, N.J. and Hogeweg, P. 1997. Modeling morphogenesis: from single cells to crawling slugs. J Theor Biol 184, 229-235.

\item
Schuster, P. 1996. How does complexity arise in evolution? Complexity 2, 22-30.

\item
Shapiro, J.A. 1998. Thinking about bacterial populations as multicellular organisms. 
Annu. Rev. Microbiol. 52, 81-104.

\item
Sims, K. 1994. Evolving 3D morphology and behavior by competition. Artif. Life 1, 353-372.

\item
Sipper, M. 1995. Using artificial life using a simple, general cellular model. J. Artif. Life 2, 1-35.

\item
Sol\'e, R.V. 2009. Evolution and self-assembly of protocells. Int. J. Biochem. Cell. Biol. 41, 274-284.

\item
Sol\'e, R.V., Montoya, J.M. and Erwin, D.H. 2002. Recovery after mass extinction: evolutionary assembly
in large-scale biosphere dynamics. Phil.  Trans. Roy. Soc. London B 357, 697-707. 

\item
Sol\'e, R.V. , Fernandez, P. and Kauffman, S.A. 2003. Adaptive walks in a gene network model of morphogenesis: insights into the Cambrian explosion. Int. J. Dev. Biol. 47: 685 - 693.  

\item
Sol\'e, R.V. , Munteanu, A., Rodriguez-Caso, C. and Macia, J. 2007. Synthetic protocell biology: from reproduction to computation.  
Phil. Trans. R. Soc. B 362, 1727-1739.

\item
Sol\'e, R.V. and Valverde, S. 2013. Before the endless forms: embodied model of transition from single cells to aggregates to ecosystem engineering. PLoS One, 8(4):e59664. doi: 10.1371/journal.pone.0059664.

\item
Szathm\'ary, E. 1994. Toy models for simple forms of multicellularity, soma and germ. J. Theor. Biol. 169, 125-132.

\item
Tusscher, K.H.  and Hogeweg, P. 2011. Evolution of networks for body plan patterning; interplay of modularity, robustness and evolvability. PLoS Comput Biol. 7, e1002208. 

\item
Yedid, G., Stredwick, J., Ofria, C.A. and Agapow, P.M. 2012. A comparison of the effects of random and selective mass extinctions on erosion of evolutionary history in communities of digital organisms. PLoS ONE 7, e37233.

\item
Wolpert, L. 1969. Positional information and the spatial pattern of cellular differentiation. J. Theor. Biol. 25, 1-47.

\end{enumerate}

\end{document}